\documentclass[12pt]{article}
\usepackage{a4}
\usepackage{amsfonts,amssymb,amsmath}
\usepackage{graphicx}
\usepackage{cite}
\usepackage{ytableau}
\newtheorem{theorem}{Theorem}

\newtheorem{proposition}[theorem]{Proposition}

\renewcommand{\i}{{\mathrm{i}}}

\begin{document}

\title{Nonassociative Weyl star products }
\author{V. G. Kupriyanov\thanks{Also at Tomsk State University, Tomsk, Russia. E.mail: vladislav.kupriyanov@gmail.com} and D. V. Vassilevich\thanks{E.mail: dvassil@gmail.com}\\
{\it CMCC-Universidade Federal do ABC, Santo Andr\'e, SP, Brazil}}

\maketitle

\begin{abstract}
Deformation quantization is a formal deformation of the algebra of smooth functions on some manifold. In the classical setting, the Poisson bracket serves as an initial conditions, while the associativity allows to proceed to higher orders. Some applications to string theory require deformation in the direction of a quasi-Poisson bracket (that does not satisfy the Jacobi identity). This initial condition is incompatible with associativity, it is quite unclear which restrictions can be imposed on the deformation. We show that for any quasi-Poisson bracket the deformation quantization exists and is essentially unique if one requires (weak) hermiticity and the Weyl condition. We also
propose an iterative procedure that allows one to compute the star product up to any desired order.
\end{abstract}

\section{Introduction}\label{sec-intro}
Star product are usually understood as deformations of the usual point-wise product
on the space of smooth functions on some manifold $M$. The theory of star products was
developed mostly in the framework of the deformation quantization approach \cite{BFFLS}.
The associativity of star products reflects the associativity of compositions of quantum
mechanical operators. A milestone result in this direction is the Formality Theorem by
Kontsevich \cite{Kontsevich} that provides a classification of star products on Poisson
manifolds and gives a recipe to calculate these products. Some more references and a historic
overview may be found in \cite{DS,Waldmann}.

The interest to star products was boosted by the discovery \cite{ChuHo,Schom} that the coordinates
of string endpoints attached to a Dirichlet brane in a background $B$-field do not commute, and
therefore, star products are natural to describe correlation functions. It was found a bit later
\cite{CoSch,MK1,MK2}, that in a non-constant $B$-field, the product should be not only
noncommutative, but also nonassociative. More recently, similar effects were discovered in
the closed string sector. It was demonstrated \cite{Lust1,Lust2,Lust3} that the presence of a
non-geometric $R$-flux leads to a twist of the Poisson structure and to nonasociativity of
corresponding star products. Note, that the mechanism of breaking the associativity is very
similar to the one that was found in the background of a magnetic monopole \cite{Jackiw}.
An extended discussion on the relations between the magnetic field and the nonassociativity may be found
in \cite{BaLu}. Despite this renewed interest, the overall situation with twisted
Poisson structures and nonassociative star products remains by far less clear than with their
associative counterparts.

We are interested in quantizing the \emph{quasi-Poisson} bracket
\begin{equation}
\{ f,g\}_Q=P^{ij}(x) \partial_i f \ \partial_j g \label{quasiP}
\end{equation}
without imposing the Jacobi identity on the bivector $P$. It is assumed usually that the failure
of $P$ to satisfy the Jacobi identity is proportional to a closed 3-form, i.e., that one deals with a
twisted Poisson structure \cite{KlSt,SWe}. We shall not use this assumption thus keeping $P$ arbitrary.
There exists a different definition of quasi-Poisson manifolds \cite{Alekseev}. We hope this will not
lead to confusion.

Star products admit a formal expansion,
\begin{equation}
f\star g = f \cdot g +\sum_{r=1}^\infty (\i\alpha)^r C_r(f,g) \,,\label{starex}
\end{equation}
where $\alpha$ is a deformation parameter, $\i\equiv \sqrt{-1}$, and $C_r$ denote some bidifferential operators. These products
provide a quantization of the bracket (\ref{quasiP}) if
\begin{equation}
C_1(f,g)-C_1(g,f)= 2\{ f,g\}_Q\,.\label{quant1}
\end{equation}
Normally \cite{BFFLS}, higher orders $C_r(f,g)$, $r >1$ , are restricted by the associativity of 
star products. The Jacobi identities on $P^{ij}$ follow from the associativity.
Any formal Poisson structure defines then a star product up to a "gauge transformation"
\cite{Kontsevich}.

In the \emph{non-Poisson/nonassociative}\footnote{There is a completely different point
of view on the quantization problem related to twisted Poisson structures. Instead of
nonassociative deformations of algebras one deals with deformations of gerbes or
algebroid stacks, see \cite{Se,ABJS,Pavlik1,Pavlik2}. With some additional restrictions, e.g. when the
twisted Poisson structure is \emph{prequantizable}, one can apply the methods of geometric quantization \cite{Petalidou}.
We shall not follow any of these approaches here.}
case the condition (\ref{quant1}) can still be imposed, but
it is unclear what may play the role of the associativity conditions in restricting the higher
orders in (\ref{starex}). There are various proposals in the literature. One can use quasi-Hopf
twist deformations \cite{MSS1}, or exploit the formality theorem and impose the Kontsevich formula
for $C_r$ as in \cite{MSS2}.

In this paper we propose an approach to nonassociative star products that is based on
realizations of coordinates $x$ as differential operators $\hat x$ and on the Weyl ordering
prescription for the functions $\hat f(\hat x)$. This approach is closer to the ideology
of quantum mechanics. One may think naively that associativity
of the operator composition contradicts nonassociativity of the star product and makes the
operator formalism useless in the present context. Constructing a nonassociative Weyl star
product  seems to be an interesting and challenging problem. We shall be able to construct
this product  for a completely general bivector $P^{ij}$. Besides, as we shall see below,
the Weyl products have considerable computations advantages over other products.

Our method for the construction of nonassociative star products generalizes the one that
we proposed earlier \cite{KV} for the associative products. The Weyl ordering is a classical
tool in the theory of star products \cite{Waldmann}. It was used to analyze Fedosov star
products on cotangent bundles \cite{BNW} and gauge theories on noncommutative space
\cite{BeSy}. Representation through differential operators also proved very convenient
\cite{BNW}, especially with the Lie-algebra type noncommutativity \cite{Gutt,Dito,Mel1,Mel2,Mel3,CO,Oriti,KV15}.

This paper is organized as follows. In the next section, we define Weyl, triangular and Hermitian
star products and formulate our main result concerning their existence and uniqueness. In Sec.\ref{sec-AJ} we derive
some properties of associator and jacobiator of star products involving coordinates. Sec.\ \ref{sec-res} contains an iterative
procedure that allows to compute the star product to any given order. Explicit expressions up to the third order are derived
in Sec.\ \ref{sec-exp}. The results are discussed in Sec. \ref{sec-disc}, while a long and important expansion can be found in the
Appendix.

\section{Main definitions and the main result}\label{sec-main}
Let $A$ be some algebra of functions on $\mathbb{R}^N$ with a point-wise product which is closed with respect to derivatives of an
arbitrary order. Consistent  choices are algebras of smooth functions or algebras of polynomials. Let $P^{ij}$ be some bivector on
$\mathbb{R}^N$, and let $A[[\alpha]]$ be the algebra of formal power series of $\alpha$ with coefficients in $A$. Star product is any
deformation of the point-wise product satisfying (\ref{starex}) and (\ref{quant1}), though we shall consider only those deformations that do
not change the unity,
  \begin{equation}\label{su}
  f\star 1=1\star f=f.
\end{equation}

Through the star product one may associate a (formal) differential operator $\hat f$ to a function $f$ as
\begin{equation}
(f\star g)(x)=\hat f \triangleright g(x)~,
\label{d3}
\end{equation}%
where the symbol $\triangleright$ on the right hand side means an action of a differential operator on
a function. In particular,
\begin{equation*}
x^i\star f=\hat x^i\triangleright f(x)~,
\end{equation*}%
We call the start product the \emph{Weyl star product} if for any $f$ the operator $\hat f$
can be obtained by the Weyl symmetric ordering of operators $\hat x^j$
corresponding to the coordinates $x^j$. If $\tilde f(p)$ is a Fourier transform of $f$, then
\begin{equation}
\hat f=\hat f\left( \hat{x}\right) =W\left( f\right)  =\int \frac{d^{N}p}{\left( 2\pi \right) ^{N}}%
\tilde{f}\left( p\right) e^{-\i p_{m}\hat{x}^{m}}.  \label{2}
\end{equation}
For example, $W(x^ix^j)=\tfrac 12 (\hat x^i\hat x^j + \hat x^j \hat x^i)$.
Weyl star products satisfy
   \begin{equation}\label{weyl}
    (x^{i_1}\dots x^{i_n})\star f=\sum_{P_n} \frac 1{n!} P_n( x^{i_1}\star(\dots \star (x^{i_n}\star f)\dots)\,.
\end{equation}
where $P_n$ denotes a permutation of $n$ elements. This equation may be even used to define Weyl star products with explicit use
of differential operators in the spirit of \cite{Dito}. For us, the use of operator language will be essential. One can write
the formal differential operator $\hat x^i$, corresponding to the coordinate $x^i$, as
\begin{equation}
\hat{x}^{i}=x^{i}+\overset{\infty }{\underset{n=1}{\sum }}\Gamma ^{i\left(
n\right) }\left( \alpha ,x\right) \left(\i \alpha \partial \right) ^{n}~,
\label{3}
\end{equation}%
where
\begin{equation}
\Gamma ^{i\left( n\right) }\left( \alpha ,x\right) =\Gamma
^{ii_{1}...i_{n}}\left( \alpha ,x\right)
\end{equation}%
We have shown in \cite{KV} that a Weyl star product satisfies the condition (\ref{su}) of stability of unity if and only if
\begin{equation}
p_{i}p_{i_{1}}...p_{i_{k}}\Gamma ^{ii_{1}...i_{k}}=0~,  \label{trless}
\end{equation}
where all indices in $\Gamma$ are contracted with commuting variables $p_i$. By the construction, Eq.\ (\ref{3}),
$\Gamma ^{ii_{1}...i_{k}}$ is antisymmetric in the last $n$ indices. Thus, $\Gamma ^{ii_{1}...i_{k}}$ may transform under the permutations
according to the representations described by the Young tableaus
\begin{equation}
\ytableausetup{mathmode}
\begin{ytableau}
i& j_1 & j_2 & \none[\dots] & j_n
\end{ytableau}
\quad \oplus \quad
\ytableausetup{mathmode}
\begin{ytableau}
j_1 & j_2 & \none[\dots] & j_n \\
i \label{YGamma}
\end{ytableau}
\end{equation}
The condition (\ref{trless}) kills the first (totally symmetric) component. Following \cite{KV} we shall call this condition the
tracelessness condition, though, strictly speaking, it is not related to any trace.

A word of warning is in order. The correspondence $f\to \hat f$ is not an algebra representation. Since the star product that we
consider here does not need to be associative, in general $\widehat{ (f\star g)} \ne \hat f \circ \hat g$.

The parameter $\alpha$ is a formal expansion parameter that has no numerical value. Therefore, in the formal setting there is no
way to compare different powers of $\alpha$, in contrast to the physical situation. For this reason, one may change within certain limits
the order assignments for various terms without changing the physical content. Alternatively, one may impose restrictions on the way
we organize the formal expansion. One possible restriction is to request a triangular structure of the expansion (\ref{starex}): the order
of derivatives acting on $f$ or $g$ in any $C_r$ may not exceed $r$. Then each $\Gamma$ is expanded in non-negative powers of $\alpha$:
\begin{equation}
\Gamma ^{i\left( n\right) }\left( \alpha ,x\right) =\overset{\infty }{%
\underset{k=0}{\sum }}(\i \alpha) ^{k}\Gamma _{k}^{i\left( n\right) }\left(
x\right) ~.  \label{4}
\end{equation}
We also impose restrictions on the terms with lowest number of derivatives. We assume that the $\Gamma^{i(n)}$ with $n=1$ has no
$\alpha$-corrections, i.e.,
\begin{equation}
\Gamma^{jk}(\alpha,x)=\Gamma^{jk}_0(x) \,.\label{Gamzer1}
\end{equation}
The meaning of this condition will be discussed later, see Sec.\ \ref{sec-disc}.
The triangular star products satisfying (\ref{Gamzer1}) will be called
\emph{strictly triangular}.

Further conditions refer to the properties of star product related with complex conjugation. A star product is called \emph{Hermitian}
if
\begin{equation}
(g\star f)^\ast=f^\ast \star g^\ast . \label{Her}
\end{equation}
We shall use a weaker condition, that we shall call \emph{weak Hermiticity}:
  \begin{equation}\label{wher}
    (x^j\star f)^\ast=f^\ast \star x^j
\end{equation}
for all $x^j$.

We are ready now to formulate the main result of this paper:

\begin{proposition}
For any bivector field $P^{ij}$ there is unique weakly Hermitian strictly triangular Weyl star product satisfying the stability of
unity condition.
\end{proposition}

Moreover, we shall present recursion relations that allow to compute this star product to any given order.

\section{Associator and Jacobiator}\label{sec-AJ}

In this section we study algebraic properties of weakly Hermitian Weyl star products.
The associator $A(f,g,h)$ and Jacobiator $J(f,g,h)$ are defined by the formulas
\begin{equation}\label{ass}
  A(f,g,h)=f\star(g\star h)-(f\star g)\star h,
\end{equation}
and
\begin{equation}\label{jac}
  J(f,g,h)=\frac{1}{6}\left\{[f,[g,h]_\star]_\star+[h,[f,g]_\star]_\star+[g,[h,f]_\star]_\star\right\},
\end{equation}
where $[f,g]_\star$ stands for the star commutator. By the definition, the Jacobiator is antisymmetric over all arguments, while the associator is not necessarily so. One finds the relation
\begin{equation}\label{jass}
   J(f,g,h)=\frac{1}{6}\left\{A(f,g,h)-A(f,h,g)+A(h,f,g)-A(h,g,f)+A(g,h,f)-A(g,f,h)\right\}.
\end{equation}
A nonassociative star product may have a vanishing Jacobiator.

For the Weyl star product, one has by (\ref{weyl})
\begin{equation}\label{xxx}
  (x^ix^j)\star x^k=\frac{1}{2}\left(x^i\star(x^j\star x^k)+x^j\star(x^i\star x^k)\right).
\end{equation}
On the other hand the stability of the unity,
\begin{equation*}
  \frac{1}{2}\left(x^i\star x^j+x^j\star x^i\right)=x^ix^j,
\end{equation*}
implies
\begin{equation}\label{j1}
 \left(x^i\star x^j\right)\star x^k+\left(x^j\star x^i\right)\star x^k=  x^i\star(x^j\star x^k)+x^j\star(x^i\star x^k).
\end{equation}
Which means that
\begin{equation}\label{j2}
  A(x^i,x^j,x^k)+ A(x^j, x^i,x^k)=0,
\end{equation}
that is, the associator of coordinate functions $x^i$, $x^j$ and $x^k$ is antisymmetric in first two arguments.

If the Weyl star product is weak-Hermitian, then considering complex conjugate of the equation (\ref{j1}) and using (\ref{wher}) we obtain
\begin{equation}\label{j3}
  x^k\star(x^j\star x^i)+x^k\star(x^i\star x^j)= \left(x^k\star x^j\right)\star x^i+\left(x^k\star x^i\right)\star x^j.
\end{equation}
That is, the associator of coordinate functions $x^k$, $x^j$ and $x^i$ is antisymmetric in the last two arguments,
\begin{equation}\label{j4}
  A(x^k,x^j,x^i)+ A(x^k, x^i,x^j)=0.
\end{equation}
Using (\ref{j2}) and (\ref{j4}) in (\ref{jass}) we find
\begin{equation}\label{j5}
  J(x^i,x^j,x^k)=A(x^i,x^j,x^k).
\end{equation}
For the weak-Hermitian Weyl star product the associator $A(x^i,x^j,x^k)$ is antisymmetric in all arguments and is equal to the Jacobiator $J(x^i,x^j,x^k)$.  Antisymmetry of the associator plays an important role in the construction of\cite{BBBS}.

%%%%%%%%%%%%%%%
\section{Resolving the weak hermiticity condition}\label{sec-res}
In this Section we shall rewrite the weak hermiticity condition as a system of coupled algebraic equations and develop an iterative
procedure that always gives a unique solution to that equations.
By Eq.\ \ref{2}, for any Weyl star product this condition is equivalent to the following equation
\begin{equation*}
   \left( \int \frac{d^{N}p }{\left( 2\pi \right) ^{N}}%
\tilde{f}\left( p\right)e^{-\i p_{j}\hat {x}^{j}}\triangleright{ x^i}\right)^\ast=\int \frac{  d^{N}p}{\left( 2\pi \right) ^{N}}%
\tilde{f}^\ast\left( p\right)\hat x^i\triangleright e^{\i p_{j}{x}^{j}}.
\end{equation*}
which holds true if and only if
\begin{equation}\label{hexp}
e^{-\i p_{j}\hat{x}^{j}}\triangleright x^i=\left( { \hat x^i}\triangleright  e^{\i p_{j}{x}^{j}}\right)^\ast.
\end{equation}
The right-hand side of this equation is
\begin{eqnarray}
  &&\left[ \left(x^{i}+\overset{\infty }{\underset{n=1}{\sum }}\Gamma ^{i\left(
n\right) }\left( \alpha ,x\right) \left(\i\alpha \partial \right) ^{n}\right)  e^{\i p_{j}{x}^{j}}\right]^\ast\label{ls}\\
&&\qquad= e^{-\i p_{j}{x}^{j}} \left(x^{i}+\overset{\infty }{\underset{n=1}{\sum }}\alpha^n(-1)^n \overset{n-1 }{\underset{k=0}{\sum }}\i^k\Gamma_k ^{i\left(
n-k\right)\ast }\left( x\right) \left(  p \right) ^{n-k}\right).\notag
\end{eqnarray}

On the left hand side of (\ref{hexp}) one needs to expand the operator $\exp[- \i p_{j}\hat{x}^{j}]$ to a given order of $%
\alpha$. To this end we shall use the Duhamel formula
\begin{equation}
e^{A+B}=e^{A}+\overset{1}{\underset{0}{\int }}e^{\left( A+B\right)
s}Be^{\left( 1-s\right) A}ds~.  \label{8}
\end{equation}%
Here $A+B=-\i p_{i}\hat{x}^{i},\ A=-\i p_{i}x^{i}$ and $B=-\i p_{i}\left( \hat{x}
^{i}-x^{i}\right)$. By using these
rules, one finds,
\begin{eqnarray}
e^{A+B} &=&e^{A}\left(1 +B+\frac{1}{2}\left[ B,A\right] +\frac{1}{2}B^{2}
\right.  \label{dec} \\
&&+\frac{1}{6}\left[ \left[ B,A\right] ,A\right] +\frac{1}{3}\left[ B,A%
\right] B+\frac{1}{6}B\left[ B,A\right] +\frac{1}{6}B^{3}  \notag \\
&&+\frac{1}{24}\left[ \left[ \left[ B,A\right] ,A\right] ,A\right] +\frac{1}{%
8}\left[ \left[ B,A\right] ,A\right] B+\frac{1}{8}\left[ B,A\right] ^{2}
\notag \\
&&+\frac{1}{24}B\left[ \left[ B,A\right] ,A\right] +\frac{1}{8} \left[ B,A%
\right] B^{2}+\frac{1}{12}B\left[ B,A\right] B  \notag \\
&&\left. +\frac{1}{24}B^{2}\left[ B,A\right] +\frac{1}{24}B^{4}+\dots \right)
 ~.  \notag
\end{eqnarray}
Let us separate on the right hand side of (\ref{dec}) the terms that are linear in $B$. These are repeated commutators of the
form $[\dots [B,A],\dots, A]$.
Denote $B_m=-\i p_j\Gamma^{j(m)}(\alpha, x)(\i\alpha\partial)^m$.
Then nonvanishing contributions may be calculated taking into account the condition (\ref{trless}):
\begin{eqnarray*}
&&[B_2,A]\triangleright x^i=2(\i\alpha)^2(-\i p_j)(-\i p_{i_1}) \Gamma^{ji_1i_2i},   \\
&&[[B_3,A],A]\triangleright x^i=3!(\i\alpha)^3(-\i p_j)(-\i p_{i_1})(-\i p_{i_2}) \Gamma^{ji_1i_2i},  \\
&&[[[B_4,A],A],A] \triangleright x^i =4!(\i\alpha)^4(-\i p_j)(-\i p_{i_1})(-\i p_{i_2})(-\i p_{i_3})
\Gamma^{ji_1i_2i_3i}.
\end{eqnarray*}
That is,
\begin{equation}\label{AB3}
  [\dots[B,\underbrace{A]\dots,A}_{m-1}]\triangleright x^i  =m!(\i\alpha )^m\Gamma^{(m) i}(\alpha, x)(-\i p)^m.
\end{equation}
Finally we conclude that the contribution of the repeated commutators to the l.h.s. of (\ref{hexp}) is given by
\begin{eqnarray}\label{ab4}
  &&e^{-\i p_{j}{x}^{j}} \overset{\infty }{\underset{m=2}{\sum }}\frac{1}{(m)!} [\dots[B,\underbrace{A]\dots,A}_{m-1}]\triangleright x^i=\\
  &&e^{-\i p_{j}{x}^{j}}\overset{\infty }{\underset{m=2}{\sum }}\alpha ^m\left(\overset{\infty }{\underset{k=0}{\sum }}(\i\alpha )^{k}\Gamma _{k}^{\left( m\right) i}\left(
x\right)\right)( p)^m=e^{-\i p_{j}{x}^{j}}\overset{\infty }{\underset{n=2}{\sum }}\alpha^n \overset{n-1 }{\underset{k=0}{\sum }}\i^k\Gamma_k ^{\left(
n-k\right)i }\left( x\right) \left(  p \right) ^{n-k}\notag
\end{eqnarray}

This allows us to rewrite the equation (\ref{hexp}) as
\begin{eqnarray}
&&\overset{\infty }{\underset{n=2}{\sum }}\alpha^n \overset{n-1 }{\underset{k=0}{\sum }}\i^k\left((-1)^n\Gamma_k ^{i\left(
n-k\right)\ast }-\Gamma_k ^{\left(
n-k\right)i }\right) \left(  p \right) ^{(n-k)}=\label{cond}\\
&&\left(\frac{1}{2}B^{2} +\frac{1}{3}\left[ B,A%
\right] B+\frac{1}{6}B\left[ B,A\right] +\frac{1}{6}B^{3}+\frac{1}{%
8}\left[ \left[ B,A\right] ,A\right] B+\frac{1}{8}\left[ B,A\right] ^{2}
\right.  \notag  \\
&&
+\frac{1}{24}B\left[ \left[ B,A\right] ,A\right] +\frac{1}{8} \left[ B,A%
\right] B^{2}+\frac{1}{12}B\left[ B,A\right] B  \notag \\
&&\left. +\frac{1}{24}B^{2}\left[ B,A\right] +\frac{1}{24}B^{4}+\dots\right)\triangleright x^i
 ~.  \notag
\end{eqnarray}
Let us also expand the right hand side of equation above in a power series of $\alpha$ and $p$
\begin{equation}
\overset{\infty }{\underset{n=2}{\sum }}\alpha^n \overset{n-1 }{\underset{k=0}{\sum }}\i^k F^{i(n-k)}(p)^{(n-k)}\,.
\label{defF}
\end{equation}

By comparing the terms with the same power of $\alpha$ and $p$ in the right and left hand sides of (\ref{cond}) we arrive at
\begin{equation}\label{7c}
  \left(\Gamma_k^{i(m)\ast }- (-1)^{k+m}\Gamma_k^{(m)i}\right)(p)^{m}=(-1)^{k+m} F_k^{i(m)}(p)^{m},\,\,\,k+m=n.
\end{equation}
With the help of the condition (\ref{trless}) and by using the symmetry properties of $\Gamma$, one derives that
\begin{equation}
\Gamma_k^{(m)i}(p)^{m} = -\frac 1m \Gamma_k^{i(m)}(p)^{m}\,.\label{symG}
\end{equation}
This yields for $m\ge 2$
\begin{eqnarray}
&&\mathrm {Re}\, \Gamma_k^{i(m)}(p)^{m} =\frac m{1+(-1)^{m+k}m} \, \mathrm{Re} F_k^{i(m)}(p)^{m}\,,\nonumber\\
&&\mathrm {Im}\, \Gamma_k^{i(m)}(p)^{m} =\frac m{1-(-1)^{m+k}m} \, \mathrm{Im} F_k^{i(m)}(p)^{m}\,.\label{ReIm}
\end{eqnarray}

Both sides of both equations (\ref{ReIm}) correspond to the same order $\alpha^n=\alpha^{m+k}$. However, since all terms on the right hand side of Eq.\ (\ref{cond}) are at least quadratic in $B$ (and, consequently, also in $\Gamma$) the tensors $F^{i(m)}_k$ are defined through the $\Gamma$'s in the order strictly less than $n$. In other words, if one knows all $\Gamma$ at the orders less than $n$, one can calculate $F^{i(m)}_k$ for $m+k=n$.

To show that the right hand sides of (\ref{ReIm}) always define uniquely the left hand sides, it is enough to show that $F^{i(m)}_k$ is always traceless. Let us pick up a term in (\ref{cond}). By the arguments similar to that around Eq.\ (\ref{AB3}), a non-vanishing contribution
of such a term should read
\begin{equation}\label{D1}
 D_{n-k-l}(\Gamma,\partial,p)\Gamma^{(l)i}_k(-\i p)^l, \,\,\,k+l<n,
\end{equation}
where we have separated explicitly the rightmost $\Gamma$. $D_{n-k-l}(\Gamma,\partial,p)$ is some differential operator. Since all $\Gamma$'s are supposed to be traceless, contracting (\ref{D1}) with $p_i$ yields zero, which means that $F^{i(m)}_k$ is traceless as well.

We see, that indeed the equations allow us to define all terms $\Gamma^{i(m)}_k$ with $m+k=n$ if the terms with $m+k<n$ are known. This procedure works for $m>1$ only, so that the case $m=1$ has to be treated separately. Due to (\ref{4}) one has to analyze $\Gamma^{ij}_0$ only.
The symmetric part vanishes because of (\ref{trless}),
\begin{equation*}
  \Gamma _{0}^{j_1i}+\Gamma _{0}^{ij_1}=0.\label{sym}
\end{equation*}
The antisymmetric part is fixed by (\ref{quant1}). Indeed, by taking $f=x^i$, $g=x^j$ we immediately obtain
\begin{equation}
\Gamma^{ij}(\alpha,x)=\Gamma _{0}^{ij}\left( x\right) =P ^{ij}\left( x\right) ~.
\label{Gamzer}
\end{equation}

Next we note that the right hand side of (\ref{7c}) does not contain imaginary coefficients. All imaginary units appear in combinations with
$\alpha$ or with $p$, but the corresponding powers of $\i$ are explicitly isolated in (\ref{defF}). Therefor, if all $\Gamma^{i(m)}_k$ with
$m+k<n$ are real, then $F^{i(m)}_k$ with $k+m=n$ are also real, and by (\ref{ReIm}) all $\Gamma^{i(m)}_k$ with $k+m=n$ are real as well.
Since the lowest order $\Gamma_0^{ij}$ is real by Eq.\ (\ref{Gamzer}), all  $\Gamma^{i(m)}_k$ are real for all $k$ and $m$. Consequently, the
only iteration relation that we need is
\begin{equation}
\Gamma_k^{i(m)}(p)^{m} =\frac m{1+(-1)^{m+k}m} \, F_k^{i(m)}(p)^{m}\,.\label{iter}
\end{equation}

An alternative way to obtain the equations (\ref{7c}) is to consider the weak hermiticity conditions taking the function $f$ in the form of
monomials,
\begin{equation}\label{7a}
 s_i x^i\star(q_{j_1}\dots q_{j_n}x^{j_1}\dots x^{j_n})=[(q_{j_1}\dots q_{j_n}x^{j_1}\dots x^{j_n})\star x^i s_i]^\ast\,.
\end{equation}
The interested reader may supply necessary technical details.

\section{Explicit formulas at lowest orders}\label{sec-exp}
\subsection{Second order}
Let us use the expansion (\ref{ExAp}) to write down the weak hermiticity condition
(\ref{hexp}) to the second order in $\alpha$:
\begin{eqnarray}
% \nonumber to remove numbering (before each equation)
  && e^{-\i p_{j}{x}^{j}}\left[x^i+\alpha p_{j_1}\Gamma _{0}^{j_1i}+\alpha^2p_{j_1}p_{j_2}\Gamma _{0}^{j_1j_2 i}+\frac{\alpha^2}{2}p_{j_1}p_{j_2}\Gamma _{0}^{j_1l_1}\partial_{l_1}\Gamma _{0}^{j_2i}\right]\label{5}\\
  &&=e^{-\i p_{j}{x}^{j}}\left[x^i-\alpha p_{j_1}\Gamma _{0}^{ij_1}+\alpha^2p_{j_1}p_{j_2}\Gamma _{0}^{ij_1j_2 }\right]+o(\alpha^2).\notag
\end{eqnarray}
Due to (\ref{sym}), there is no new restriction at the order of $\alpha^1$, though
at the second order we obtain the algebraic equation on $\Gamma_0^{ijk}$ in terms of $\Gamma_0^{ij}$:
\begin{equation}\label{Gam2}
(\Gamma _{0}^{ij_1j_2 }-\Gamma _{0}^{j_1j_2 i})  p_{j_1}p_{j_2}=\frac{1}{2}\Gamma _{0}^{j_1l_1}\partial_{l_1}\Gamma _{0}^{j_2i} p_{j_1}p_{j_2}\,,
\end{equation}
which yields
\begin{equation}\label{7}
  \Gamma_0^{ijk}=\frac{1}{6}\left( P^{kl}\partial_l P^{ji}+P^{jl}\partial_l P^{ki}\right).
\end{equation}

The star product is then recovered by using the relation $f\star g =\hat f \triangleright g$ and the representation (\ref{2}) for the
Weyl ordered operator $\hat f$ together with the expansion (\ref{ExAp}). In short, one can take (\ref{ExAp}) and replace there all partial
derivatives by derivatives of $g$ and all momenta $(-\i q_j)$ by partial derivatives of $f$, $\partial_j f$.
Taking into account (\ref{Gamzer}), (\ref{7}) and the decomposition in the appendix the expression for the star product up to the third order reads
\begin{eqnarray}
&&(f\star g)(x)=f\cdot g+\i\alpha \partial
_{i}fP ^{ij}\partial _{j}g  \label{star2} \\
&&-\frac{\alpha ^{2}}{2}P^{ij}P ^{kl}\partial _{i}\partial
_{k}f\partial _{j}\partial _{l}g-\frac{\alpha ^{2}}{3}P ^{ij}\partial
_{j}P ^{kl}\left( \partial _{i}\partial _{k}f\partial _{l}g-\partial
_{k}f\partial _{i}\partial _{l}g\right) +o\left( \alpha ^{2}\right) ~.
\notag
\end{eqnarray}

\subsection{Third order}

In the third order we need algebraic equations on two functions $\Gamma_0^{ijkl}$ and $\Gamma_1^{ijk}$. The first one reads
\begin{equation}\label{G30}
  (\Gamma_0^{ijkl}+\Gamma_0^{jkli})q_jq_kq_l=-F_0^{ijkl}q_jq_kq_l,
\end{equation}
where
\begin{equation}\label{F30}
  F_0^{ijkl}=\frac{1}{6}P^{jm}\partial_mP^{kn}\partial_nP^{li}+\frac{1}{6}P^{jm}P^{kn}\partial_m\partial_nP^{li}
  +\frac{2}{3}\Gamma_0^{jkm}\partial_mP^{li}+\frac{1}{3}P^{jm}\partial_m\Gamma_0^{kli}\,.
\end{equation}
The second algebraic equation is
\begin{equation}\label{G31}
   (\Gamma_1^{ijk}+\Gamma_1^{jki})q_jq_k=-F_1^{ijk}q_jq_k,
\end{equation}
where
\begin{equation}\label{F31}
  F_1^{ijk}=\frac{1}{2}\Gamma_0^{kmn}\partial_m\partial_nP^{ji}.
\end{equation}
Both equations can be solved easily by using the relation (\ref{iter}). One obtains
\begin{equation}\label{Gam30}
  \Gamma_0^{ijkl}=\frac{1}{18}\left(P^{jm}P^{kn}\partial_m\partial_nP^{il}+
  P^{lm}P^{kn}\partial_m\partial_nP^{ij}+P^{jm}P^{ln}\partial_m\partial_nP^{ik}\right),
\end{equation}
and
\begin{equation}\label{Gam31}
  \Gamma_1^{ijk}=\frac{1}{6}\left(
  P^{lm}\partial_lP^{kn}\partial_m\partial_nP^{ij}+P^{lm}\partial_lP^{jn}\partial_m\partial_nP^{ik}\right).
\end{equation}

The star product can be obtained in the manner as we have already described above Eq.\ (\ref{star2}). Let us write the 3rd order part of it as
\begin{equation}
f\star _{3}g=-\i\alpha^3 \sum L_{m,n}\left( f,g\right) ~,  \label{11}
\end{equation}%
where each term in the sum has $m$ derivatives acting on $f$ and
$n$ derivatives acting on $g$. Since, $\Gamma_2^{ij}=0,$
\begin{equation}\label{L11}
  L_{1,1}\left( f,g\right)=0.
\end{equation}
Further terms read
\begin{eqnarray}\nonumber
  L_{1,2}\left( f,g\right)  &=&\partial _{i}f\Gamma _{2}^{ijk}\partial
_{j}\partial _{k}g=\frac{1}{3}\partial _{i}fP^{lm}\partial_lP^{kn}\partial_m\partial_nP^{ij}\partial
_{j}\partial _{k}g\,,
\\
  L_{1,3}\left( f,g\right)  &=&\partial _{i}f\Gamma _{1}^{ijkl}\partial
_{j}\partial _{k}\partial _{l}g=\frac{1}{6}\partial _{i}fP^{jm}P^{kn}\partial_m\partial_nP^{il}\partial
_{j}\partial _{k}\partial _{l}g~\,,\nonumber\\
  L_{2,3}\left( f,g\right)  &=& =\frac{1}{3}\partial _{i}\partial _{j}fP^{ik}P^{ln}\partial_lP^{jm}\partial_k\partial_m\partial_ng\,,\nonumber\\
L_{2,2}(f,g)&=&\partial_i\partial_jf  \left[\frac{3}{2}\Gamma_0^{ijkl}+\frac{1}{2}P^{im}\partial_m\Gamma_0^{jkl}+\Gamma_0^{ikm}\partial_mP^{jl}\right]
\partial_k\partial_lg\\
&=&  \frac{1}{6}  P^{nk}\partial_nP^{jm}\partial_mP^{il}
\left(\partial_i\partial_jf\partial_k\partial_lg-\partial_i\partial_jg\partial_k\partial_lf\right).\nonumber
\end{eqnarray}
One can check at this order that
\begin{equation}\label{11111}
  L_{m,n}(f,g)=-L_{n,m}(g,f).
\end{equation}

By collecting all the terms above, we obtain
\begin{eqnarray}
f \star_3 g&=
&-\i\alpha^3 \left[
\frac{1}{3}P ^{nl}\partial _{l}P ^{mk}\partial _{n}\partial
_{m}P ^{ij}\left( \partial _{i}f\partial _{j}\partial _{k}g-\partial
_{i}g\partial _{j}\partial _{k}f\right) \right.  \notag \\
&&+ \frac{1}{6}  P^{nk}\partial_nP^{jm}\partial_mP^{il}
\left(\partial_i\partial_jf\partial_k\partial_lg-\partial_i\partial_jg\partial_k\partial_lf\right)   \notag \\
&&+\frac{1}{3}P ^{ln}\partial _{l}P ^{jm}P ^{ik}\left(
\partial _{i}\partial _{j}f\partial _{k}\partial _{n}\partial _{m}g-\partial
_{i}\partial _{j}g\partial _{k}\partial _{n}\partial _{m}f\right)
\label{16} \\
&&+\frac{1}{6}P ^{jl}P ^{im}P ^{kn}\partial _{i}\partial
_{j}\partial _{k}f\partial _{l}\partial _{n}\partial _{m}g  \notag \\
&&\left. +\frac{1}{6}P ^{nk}P ^{ml}\partial _{n}\partial _{m}P
^{ij}\left( \partial _{i}f\partial _{j}\partial _{k}\partial _{l}g-\partial
_{i}g\partial _{j}\partial _{k}\partial _{l}f\right)\right] ~.  \notag
\end{eqnarray}
We notice that the calculation of this star product is relatively easy.

%%%%%%%%%%%%%%%
\section{Discussion}\label{sec-disc}
In this paper we have demonstrated that for any bivector field, there exists a unique Weyl weakly Hermitian strictly triangular star product.
Our proof is constructive: the coefficients $\Gamma^{i(m)}_k$ are defined by (\ref{iter}) through the coefficients $F^{i(m)}_k$, which are
defined through the expansion (\ref{defF}) by lower order $\Gamma^{i(m)}_k$. By repeating these steps one can express any $\Gamma^{i(m)}_k$
through $\Gamma^{ij}$, which appears to be our bivector $P^{ij}$, see (\ref{Gamzer}). One may allow $\Gamma^{ij}$ to have higher order
contributions, but the corresponding ambiguity is rather non-interesting. Physically, one may say that the higher order corrections are
 absorbed in renormalization of the bivector $P^{ij}$. The renormalization is of course possible since $P^{ij}$ is arbitrary. This is in
contrast to associative Weyl star products. While the initial bivector in the associative case is a Poisson bivector, the corrections are
non-Poisson \cite{PenV,KV}. Consequently, the higher order corrections are non-removable and important\footnote{According to Dito
\cite{GD15} these corrections correspond to Kontsevich diagrams with wheels.}. Therefore, just imposing the Jacobi identity on $P^{ij}$ does
not make our star product associative, one still needs additional corrections terms in $\Gamma^{ij}$.

Let us study which of our requirements are satisfied by the nonassociative star products appearing in the literature 
\cite{CoSch,MK1,MK2,MSS1,MSS2,BaLu}. The triangularity condition is automatic if one likes to express the star product through the
bivector $P^{ij}$ and its derivatives only\footnote{Of all papers cited above just a single one \cite{MK2} uses an additional structure 
(the Born-Infeld measure) apart of the bivector to construct the star product. However, even the star product of Ref.\ \cite{MK2} can
be made triangular by assigning properly the orders of deformation parameter.}. Then, one does not even need to write the deformation 
parameter explicitly since the order in the perturbative expansion may be counted by the order of $P^{ij}$. Since $P^{ij}$ has two
upper indices, in $C_r(f,g)$ one has $2r$ upper indices to be contracted with lower indices of the derivatives acting on $P$, $f$ and $g$.
Since the terms containing $P^{ij}\partial_i\partial_j f$ vanish, one has at most $r$ derivatives acting on each of the functions. This means
that the star product is triangular, cf. \cite{CoSch,MK1,MSS1,MSS2,BaLu}. There is no fundamental reason to expect a start product being
\emph{strictly} triangular, though this property is important to show the uniqueness and easy to achieve by a renormalization of $P^{ij}$ (see above). The products of \cite{CoSch,MK1,MSS1,MSS2,BaLu} are strictly triangular (at least to the order to which these products are written
down explicitly), while the product of \cite{MK2} is not.

The hermiticity condition (\ref{Her}) is very natural, as well as its weaker version (\ref{wher}). Such conditions are usually imposed in
all quantization schemes. Therefore, no wonder that all products \cite{CoSch,MK1,MK2,MSS1,MSS2,BaLu} appear to be hermitian, at least to the
order to which they are known. On the contrary, the Weyl ordering reflects a very particular quantization prescription. To obtain a unique
star products one should of course fix the quantization prescription, but this is also the main reason why our star product may differ from
other star products.

The papers
\cite{CoSch,MK1} contain the whole 2nd order and some terms in the 3rd order of star product following from the open strings/D-branes calculations. All these terms are in perfect agreement with our expressions (\ref{star2}) and (\ref{16}). Probably, this is not 
accidental but postpone the complete analysis to a future work. 

Our product differs from the one
used in the context of the $R$-flux string models \cite{MSS1,MSS2,BaLu}, and also from the one computed by the Kontsevich formulas
\cite{Kontsevich}. (There is some degree of confusion in the literature regarding the form of Kontsevich product at the second order
of deformation parameter. See \cite{Dito} for explicit calculations.) All these products are not Weyl. Since our method works for a
completely general bivector $P^{ij}$ we do not have to impose any restrictions of the $R$-fluxes, distributions of magnetic charge, etc.
One may easily check that our star product reproduces the same commutators and the same jacobiators involving the coordinates as 
that of Refs.\ \cite{MSS1,MSS2,BaLu}, though other quantities may, of course, differ.

Associative star products may be represented through path integrals of Poisson sigma model \cite{SchS} on a disc \cite{CF}
or on a finite cylinder \cite{DV}. It is an open problem to verify whether our nonassociative star product also admits a path integral interpretation.

We have good grounds to believe that the full hermiticity property (\ref{Her}) holds for our star product. We were able to check this at
some examples, but a complete proof is still missing.

\section*{Acknowledgements}
We are grateful to Jim Stasheff for his detailed comments on the previous version of this manuscript.
This work was supported in part by CNPq, projects 443436/2014-2 (V.K.),
306208/2013-0 (D.V.) and 456698/2014-0 (D.V.), and FAPESP, projects 2014/03578-6 (V.K.) and 2012/00333-7 (D.V.).

\appendix
\section{Important expansion}
Here using the form (\ref{3}) of the operator $\hat x^i$ and the decomposition (\ref{dec}) we obtain the expression for the operator 
$\exp(-\i q_i\hat x^i)$ up to the third order in $\alpha$:
\begin{eqnarray}
  e^{-\i q_i\hat x^i}&=&  e^{-\i q_i x^i}\left(1+\i\alpha(-\i q_i)\Gamma_0^{ij}\partial_j\right. \nonumber\\
&&+ \alpha^2\left\{(-\i q_i)\Gamma_0^{ijk}\partial_j\partial_k+(-\i q_i)(-\i q_j)\left[\Gamma_0^{ijk}+\frac{1}{2}\Gamma_0^{im}\partial_m \Gamma_0^{jk} \right]\partial_k\right.\nonumber\\
&&+\left.(-\i q_i)(-\i q_j)\frac{1}{2}\Gamma_0^{ik}\Gamma_0^{jl}\partial_k\partial_l\right\}\label{ExAp}\\
&&-\i\alpha^3\left\{(-\i q_i)\Gamma_1^{ijk}\partial_j\partial_k+(-\i q_i)\Gamma_0^{ijkl}\partial_j\partial_k\partial_l\right.\nonumber\\
&&+(-\i q_i)(-\i q_j)\left[\Gamma_1^{ijk}+\frac{1}{2}\Gamma_0^{imn}\partial_m\partial_n\Gamma_0^{jk}\right]\partial_k
+(-\i q_i)(-\i q_j)\Gamma_0^{ik}\Gamma_0^{jlm}\partial_k\partial_l\partial_m\nonumber\\
&&+(-\i q_i)(-\i q_j)\left[\frac{3}{2}\Gamma_0^{ijkl}+\frac{1}{2}\Gamma_0^{im}\partial_m\Gamma_0^{jkl}+\Gamma_0^{ikm}\partial_m\Gamma_0^{jl}\right]\partial_k\partial_l\nonumber\\
&&+(-\i q_i)(-\i q_j)(-\i q_k)\left[\Gamma_0^{ijkl}
+\frac{1}{6}\Gamma_0^{im}\partial_m\Gamma_0^{jn}\partial_n\Gamma_0^{kl}+\frac{1}{6}\Gamma_0^{im}\Gamma_0^{jn}\partial_m\partial_n\Gamma_0^{kl}\right.\nonumber\\
&&+\left.\frac{2}{3}\Gamma_0^{ijm}\partial_m\Gamma_0^{kl}+\frac{1}{3}\Gamma_0^{im}\partial_m\Gamma_0^{jkl}\right]\partial_l\nonumber\\
&&+(-\i q_i)(-\i q_j)(-\i q_k)\left[\frac{1}{2}\Gamma_0^{im}\Gamma_0^{jn}\partial_n\Gamma_0^{kl}+\Gamma_0^{ijm}\Gamma_0^{kl}\right]\partial_l\partial_m\nonumber\\
&&+\left.(-\i q_i)(-\i q_j)(-\i q_k)\frac{1}{6}\Gamma_0^{im}\Gamma_0^{jn}\Gamma_0^{kl}\partial_l\partial_m\partial_n\right\}+o\left(\alpha^3\right).
\nonumber
\end{eqnarray}

\end{document}